\documentclass[useAMS,usenatbib,usegraphicx]{mn2e}
\usepackage{amssymb}

\title[Constraints on supermassive dark stars]{Observational constraints on supermassive dark stars}
\author[Zackrisson et al.]{Erik Zackrisson$^{1,7}$\thanks{E-mail: ez@astro.su.se}, Pat Scott$^{2,7}$, Claes-Erik Rydberg$^{1,7}$, Fabio Iocco$^{3}$, \newauthor Sofia Sivertsson$^{4,7}$, G\"oran \"Ostlin$^{1,7}$, Garrelt Mellema$^{1,7}$, Ilian T.~Iliev$^{5}$ \newauthor and Paul R.~Shapiro$^{6}$\\ 
$^{1}$Department of Astronomy, Stockholm University, 10691 Stockholm, Sweden\\
$^{2}$Department of Physics, Stockholm University, 10691 Stockholm, Sweden\\
$^{3}$Institut d'Astrophysique de Paris, UMR 7095-CNRS Paris, Universit\'e Pierre et Marie Curie, Boulevard Arago 98bis, 75014, Paris, France\\
$^{4}$Department of Theoretical Physics, Royal Institute of Technology (KTH), 10691 Stockholm, Sweden\\
$^{5}$Astronomy Centre, Department of Physics \& Astronomy, Pevensey II Building, University of Sussex, Falmer,\\ \hspace{0.25cm} Brighton BN1 8QH, United Kingdom\\
$^{6}$Department of Astronomy and Texas Cosmology Center, The University of Texas at Austin, Austin, TX 78712, USA\\
$^{7}$Oskar Klein Centre for Cosmoparticle Physics, AlbaNova University Centre, 10691 Stockholm, Sweden}

\begin{document}

\date{Accepted ... Received ...; in original form ...}

\pagerange{\pageref{firstpage}--\pageref{lastpage}} \pubyear{2010}

\maketitle

\label{firstpage}

\begin{abstract}
Some of the first stars could be cooler and more massive than standard stellar models would suggest, due to the effects of dark matter annihilation in their cores.  It has recently been argued that such objects may attain masses in the $10^4$--$10^7\ M_\odot$ range and that such supermassive dark stars should be within reach of the upcoming {\it James Webb Space Telescope}. Notwithstanding theoretical difficulties with this proposal, we argue here that some of these objects should also be readily detectable with both the {\it Hubble Space Telescope} and ground-based 8--10 m class telescopes.  Existing survey data already place strong constraints on $10^7\ M_\odot$ dark stars at $z\approx 10$.  We show that such objects must be exceedingly rare or short-lived to have avoided detection.
\end{abstract}


\section{Introduction}
\label{intro}
WIMPs (Weakly-Interacting Massive Particles) are one of the most promising candidates for dark matter \citep[DM; see e.g.][]{Jungman96, Bergstrom00}.  WIMPs may congregate and annihilate in stellar cores, affecting the appearance and evolution of their host stars \citep[e.g.][]{SalatiSilk, Moskalenko07, Spolyar08, Fairbairn08, Iocco et al. 2008, Scott09}.

WIMP annihilation in the dark matter halos that hosted galaxy and primordial star formation may have affected the ability of halo baryons to form stars there, by heating the gas enough to compete with radiative cooling \citep{Ascasibar07}.  Of particular interest is the idea that the first stars (Population III) might have been substantially affected, not only because this heating competes with the cooling necessary to form a star, but because it can lead to a new phase of stellar evolution dominated by annihilation rather than fusion \citep{Spolyar08}.  It has been suggested that this phase could be long-lived \citep[e.g.][]{Freese08}, but evidence in this direction is far from conclusive \citep{Ripamonti10, Sivertsson10}.  If they are long-lived, such stars could be detectable \citep{Zackrisson10} with the upcoming {\it James Webb Space Telescope (JWST)}, scheduled for launch in 2014.  In this scenario, the dominant effect is gravitational contraction of the host halo by baryonic infall associated with the formation of the star \citep{Spolyar08, Iocco et al. 2008, Freese09}.  Following the initial contraction, collisions of through-going WIMPs with stellar nuclei, and their subsequent scattering to lower-energy orbits, might bring yet more dark matter into the star \citep{Gould87, Iocco08, Freese08}.  

Because such dark matter-powered stars (also commonly dubbed ``dark stars'') are cooler than normal Population III stars, they are able to accrete more mass than their canonical cousins before radiative feedback develops, leading to stellar masses of up to $\sim10^3 M_\odot$.  To a first approximation, the limiting stellar mass is set only by the point at which the star can no longer be sustained by dark matter annihilation; eventually, either the dark matter is depleted by annihilation, or the star becomes too massive for the annihilation to support it.  In realistic models, one might also expect aspherical accretion, stellar winds and three-dimensional gas flows in the upper atmosphere to cause some degree of fragmentation, probably leading to somewhat smaller final masses. 

\citet{Freese et al. 2010} have recently suggested the existence of \emph{supermassive} dark stars (SMDS), with masses of up to $10^7 M_\odot$.  As is to be expected, such mega-stars would be easily detectable with {\it JWST}, even at redshift $z\approx 10$--15.

In order to achieve such masses, the authors postulate an effectively boundless well of dark matter from which a star might draw its power.  They argue that the depletion of the dark matter accessible to the growing dark star can be wholly avoided by the existence of chaotic orbits in triaxial halos, as these will refill the region of phase space shown to be depleted very quickly by \citet{Sivertsson10}.

However, the depleted WIMPs belong to a qualitatively different population than those on chaotic orbits.  The depleted population consists entirely of bound WIMPs, whose orbits lie within the star's sphere of influence.  These are the WIMPs that are gravitationally contracted into a dark star, and go on to power it.  WIMPs arriving on chaotic orbits come from an unbound population.  Members of the unbound population typically do not contribute significantly to the annihilation rate \citep{Sivertsson10}, because their occupation time in the star is extremely low (i.e.~they are moving very quickly, and do not return to pass through the star again).

It is very difficult to exchange WIMPs between the two populations.  The only way for unbound WIMPs to repopulate the depleted region of phase space is to scatter on stellar nuclei, and become gravitationally bound to the star.  Without scattering, unbound WIMPs can only significantly contribute to the annihilation rate if very many of them pass through the star per unit time.  In this case, their small individual occupation times would be offset by the sheer number passing through the star. The SMDS scenario hence requires either an extremely high rate of centre-crossing by WIMPs on chaotic orbits, or a large centre-crossing rate combined with a very large nuclear scattering cross-section.

WIMP-nucleon scattering cross-sections are strongly constrained by experiment \citep[e.g.][]{IceCube,CDMS}, requiring $\sigma \lesssim 10^{-38}$\,cm$^2$ for scattering on hydrogen.  Without detailed simulations of triaxial WIMP halos, it is difficult to comment on exactly how hard it is to achieve a high rate of centre-crossing by WIMPs on chaotic orbits.  In a spherical halo, with an isotropic distribution of WIMP velocities outside the star's sphere of influence, and assuming $\sigma = 10^{-38}\, cm^2$, the resultant conversion rate of unbound to bound WIMPs is $\sim$8 orders of magnitude too small to produce SMDS \citep{Sivertsson10}.  For SMDS to be viable, this additional factor of $10^8$ must be compensated for entirely by the radial bias of chaotic orbits.  This would seem quite improbable, though more detailed calculations are required to definitively rule it out.

\citet{Freese et al. 2010} do not explicitly calculate the rate of change of dark matter annihilation in their stars, simply assuming that the dark matter population required to support a given stellar mass is always present.  One might expect substantial differences in the evolutionary history of such stars if the dark matter density was provided self-consistently.  For example, any downwards perturbation in the dark matter annihilation rate leads to a contraction of the dark star.  This leaves behind a shell of dark matter \citep{Iocco et al. 2008}, further reducing the total annihilation rate.  The only way to avoid this is for the gravitationally-contracted dark matter to thermalise in the protostar very quickly, changing its distribution from effectively flat with radius to strongly peaked at the centre of the star.  Again, this requires a very large nuclear scattering cross-section.

\citet{Freese et al. 2010} rightly question the overall stability of SMDS, particularly with respect to general-relativistic corrections to their gravitational potentials.  Regardless of their gravitational (in)stability though, SMDS should also suffer from radiative-hydrodynamic instabilities.  As expected in radiatively-supported objects, their luminosities all lie on the Eddington limit.  This means that they probably exhibit significant continuum-driven stellar winds, not to mention line-driven ones from H and He.  The effect of including these processes in detail would almost certainly be to limit their masses to something lower than found by \citet{Freese et al. 2010}.  Any upward perturbation in the dark matter annihilation rate would substantially accelerate this effect.  The growth of SMDS hence requires a very delicately tuned, smoothly varying annihilation rate, in order for the accretion to continue as proposed by \citet{Freese et al. 2010}.

As well as theoretical problems with the SMDS formation scenario, there are also observational constraints. Here, we use stellar atmosphere models to estimate the apparent magnitudes of these dark stars. We argue that there is no need to wait for the {\it JWST} to observationally test the SMDS hypothesis, since the most massive ($\sim 10^7\ M_\odot$) SMDS are sufficiently bright to be readily detected with the Hubble Space Telescope ({\it HST}) or 8--10 m class telescopes on the ground. In fact, such SMDS would need to be exceedingly rare and/or short-lived to have evaded detection until now.

\section{The apparent magnitudes of supermassive dark stars}
\label{magnitudes}
We use the TLUSTY stellar atmosphere code \citep{Hubeny & Lanz} to generate spectra in the 0.015--300 $\mu$m wavelength range for the SMDS of \citet{Freese et al. 2010}. To derive broadband fluxes, these model spectra are redshifted to $z=0$--15 and convolved with the relevant filter transmission profiles and detector sensitivities. The resulting broadband fluxes are converted into apparent magnitudes using the luminosity distance, assuming $H_0=72$ km s$^{-1}$ Mpc$^{-1}$, $\Omega_M=0.27$ and $\Omega_\Lambda=0.73$. The calibration is based on AB magnitude system, defined so that an object with a constant flux per unit frequency interval of 3631 Jy has zero AB magnitudes $m$ in all filters. To simulate the Gunn-Peterson trough due to the opaque intergalactic medium during the reionization epoch, all fluxes are moreover set to zero at restframe wavelengths shortward of Ly$\alpha$ ($\lambda<0.1216\ \mu$m) at $z>6$ \citep{Fan et al.}.

The resulting broadband fluxes are valid at the surfaces of these objects, and neglects the fact that many of the SMDS of \citet{Freese et al. 2010} also are sufficiently hot to photoionize the surrounding gas. The resulting HII regions will add emission lines and a nebular continuum to the observed spectra of these SMDS, thereby substantially boosting their fluxes in the rest-frame UV and optical \citep[e.g.][]{Schaerer}. Because of this, the fluxes we derive should be regarded as conservative.  

\begin{figure}
\includegraphics[width=84mm]{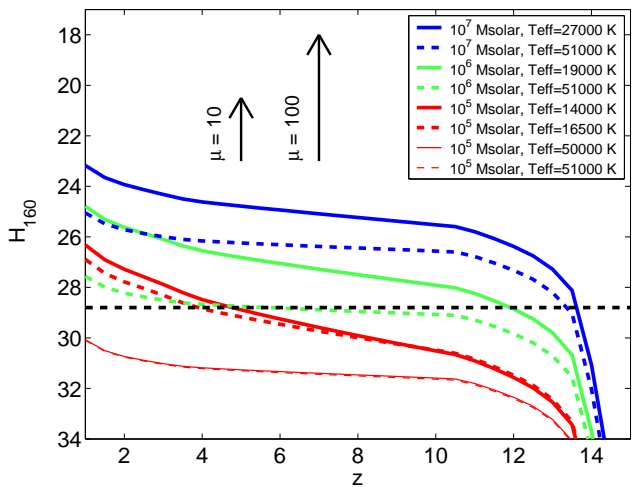}
\caption{The apparent $H_{160}$ AB magnitudes of $10^5$--$10^7\ M_\odot$ SMDS from \citet{Freese et al. 2010} as a function of redshift. The coloured lines represent dark star models with $10^7\ M_\odot$ and $T_\mathrm{eff}=27000$ K (thick solid blue), $10^7\ M_\odot$ and $T_\mathrm{eff}=51000$ K (thick dashed blue), $10^6\ M_\odot$ and $T_\mathrm{eff}=19000$ K (thick solid green), $10^6\ M_\odot$ and $T_\mathrm{eff}=51000$ K (thick dashed green), $10^5\ M_\odot$ and $T_\mathrm{eff}=14000$ K (thick solid red) and $10^5\ M_\odot$ and $T_\mathrm{eff}=16500$ K (thick dashed red). The thin solid red and thin dashed red lines (which nearly overlap in this diagram) represent the $10^5\ M_\odot$ dark stars with $T_\mathrm{eff}=50000$ K and $T_\mathrm{eff}=51000$ K, respectively. The dashed horizontal line indicates the detection threshold of the WFC3 HUDF observations \citep[e.g.][]{Bouwens et al. b}. Both $10^7\ M_\odot$ models (thick blue lines) are readily detectable above the threshold at $z=10$, and so is one of the $10^6\ M_\odot$ models (thick solid green line). The vertical lines indicate how the models would be shifted by a gravitational magnification of $\mu = 10$ (short arrow) and $\mu = 100$ (long arrow) from a foreground galaxy cluster. A lensing boost of $\mu = 10$ would lift the remaining $10^6\ M_\odot$ model (thick dashed green line) and two of the $10^5\ M_\odot$ models (thick solid and dashed red lines) into the detectable range. A boost of $\mu=100$ would make all the plotted models detectable.} 
\label{fig1}
\end{figure}

In Fig.~\ref{fig1}, we plot the redshift evolution of the apparent magnitudes for $10^5$--$10^7\ M_\odot$ SMDS in the {\it HST} F160W filter (hereafter $H_{160}$). At $z=10$, both $10^7\ M_\odot$ SMDS (thick blue solid and dashed lines) and one of the $10^6\ M_\odot$ dark stars (thick solid green line) from \citet{Freese et al. 2010}  are readily detectable above the $H_{160}=28.8$ mag threshold of current Wide Field Camera 3 (WFC3) images taken of the Hubble Ultra Deep Field (HUDF) in August 2009 ({\it HST} GO 11563: PI Illingworth). Similar observations of fields gravitationally lensed by foreground lensing clusters could in principle lift the remaining SMDS in this diagram above the detection threshold. This is illustrated by the two arrows, which indicate how the lines would be shifted by magnifications of $\mu=10$ (short arrow) and $\mu=100$ (long arrow). By comparison, a galaxy of similar mass would be considerably fainter than these SMDS. Using the \citet{Zackrisson01,Zackrisson08} population synthesis model, we predict that a galaxy with stellar mass $10^7\ M_\odot$,  metallicity $Z=0.001$ and Salpeter IMF (mass range 0.08--120$\ M_\odot$) should have $H_{160}>29$ mag at $z=10$, i.e. at least 2.5 mag fainter than a $10^7\ M_\odot$ SMDS.  

\begin{figure}
\includegraphics[width=84mm]{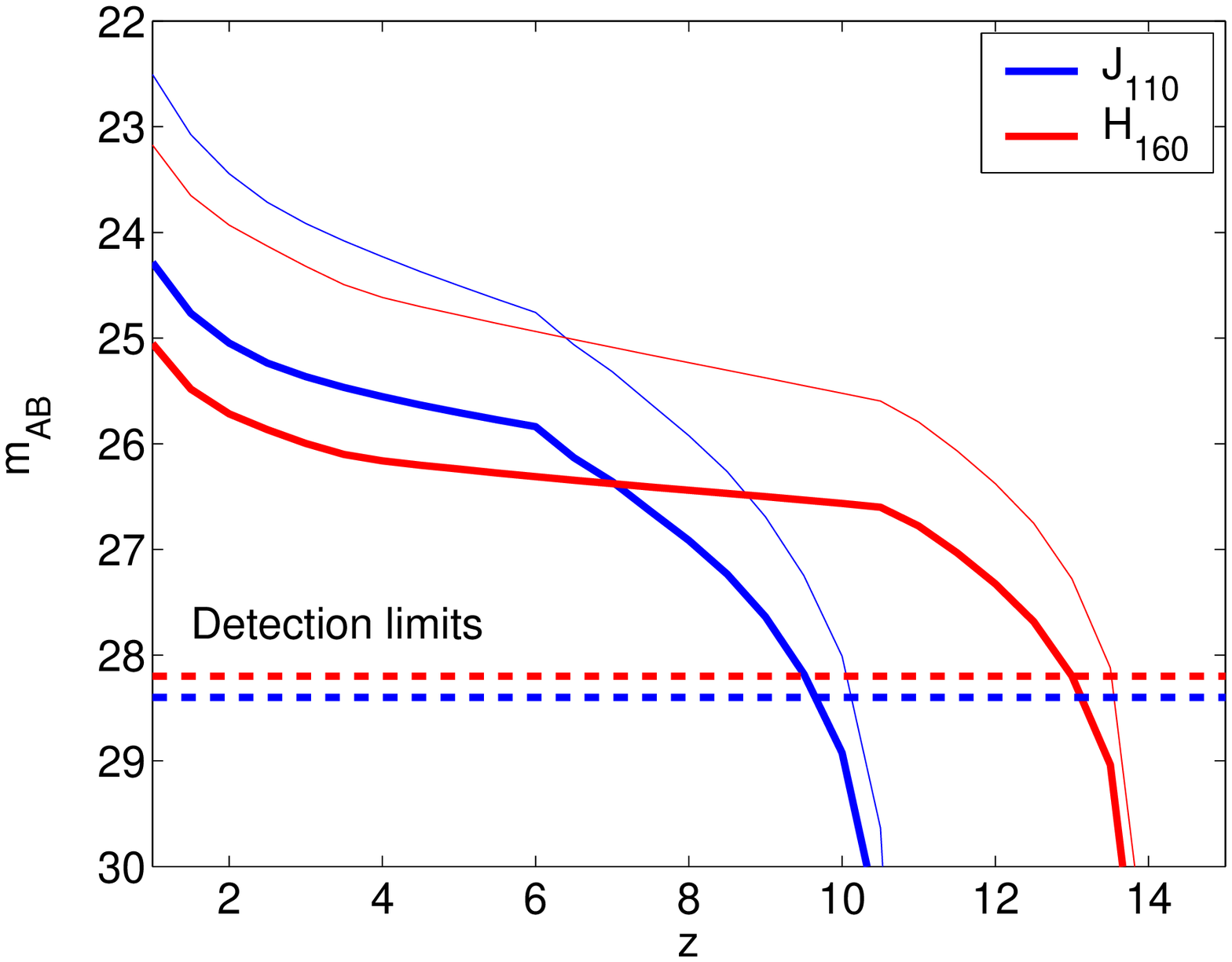}
\caption{The apparent magnitudes of $10^7\ M_\odot$ SMDS as a function of redshift. Thin lines represent the $T_\mathrm{eff}=27000$ K and the thick lines the $T_\mathrm{eff}=51000$ K SMDS from \citet{Freese et al. 2010}. The blue lines represent fluxes in the $J_{110}$ filter and red lines fluxes in the $H_{160}$ filter. The vertical dashed lines mark the detection limits of the deepest survey data compiled by \citep{Bouwens et al. a}. The rapidly dropping $J$-band fluxes at $z>6$ are due to the Gunn-Peterson trough, making such objects appear as $J$-band dropouts ($J_{110}-H_{160}>1.3$) at $z>9$. An similar decline is also seen in the $H_{160}$ band at $z>10.5$.} 
\label{fig2}
\end{figure}

To be able to separate $z=10$ objects from low-redshift interlopers, data in more than one filter is required. Since the Gunn-Peterson trough renders the flux in the F110W filter (hereafter $J_{110}$) extremely low at $z\approx10$, objects at this redshift will remain undetected in $J_{110}$ (and at all shorter wavelengths) while still being potentially detectable in $H_{160}$. Because of this effect, candidate $z\approx 10$ objects can be selected from large multiband datasets by searching for $J$-band dropouts. This is demonstrated in Fig.~\ref{fig2}, where we plot the redshift evolution of the two $10^7\ M_\odot$ SMDS from \citet{Freese et al. 2010} in both $J_{110}$ and $H_{160}$ bands. The $J_{110}$ band fluxes (blue lines) drop rapidly at $z>6$ as the Gunn-Peterson trough enters this filter, making the $z\approx 10$ fluxes extremely low (implying very red $J-H$ colours). Eventually, the Gunn-Peterson trough also starts to affect the $H_{160}$ filter, but not until $z> 10.5$.

Since the most massive SMDS should be sufficiently bright to already be readily detectable, one might ask what existing deep near-IR survey data have to say about the existence of such objects. For this purpose, we use the the $J_{110}$ and $H_{160}$ data compiled by \citet{Bouwens et al. a} from {\it HST} and groundbased images of the regions around the HUDF and the Hubble Deep Field-North. While the more recent WFC3 data reach slightly deeper than these, they have not yet been as thoroughly analysed, and the number of bona fide $J$-band dropouts in the WFC3 images still remains controversial \citep[cf.][]{Yan et al.,Bouwens et al. b}. 

The \citet{Bouwens et al. a} compilation contains a number of survey areas with slightly different detection thresholds in the $J_{110}$ and $H_{160}$ passbands, reaching a maximum depth of $J_{110}=28.4$ and $H_{160}=28.2$ at $5\sigma$ (dashed blue and red horizontal lines in Fig.~\ref{fig2}). By adopting a colour criteria of $J_{110}-H_{160}>1.3$ for $J$-band dropouts, and cross-correlating potential detections with data at other wavelengths, \citet{Bouwens et al. a} reported a null detection of credible $J$-band dropouts in these images. Still, Fig.~\ref{fig2} indicates that the two $10^7\ M_\odot$ SMDS should appear as $J$-band dropouts according to these colours criteria, at least throughout the redshift range $z\approx 9.5$--10.5. This conflict can only be resolved if $10^7 M_\odot$ SMDS are so rare or shortlived that no such objects are expected within the surveyed regions of the sky. In the following sections, we attempt to convert this criterion into a quantitative constraint on the formation rate and properties of $10^7 M_\odot$ SMDS.

\section{Upper limits on supermassive dark stars}
\label{results}
\begin{figure}
\includegraphics[width=84mm]{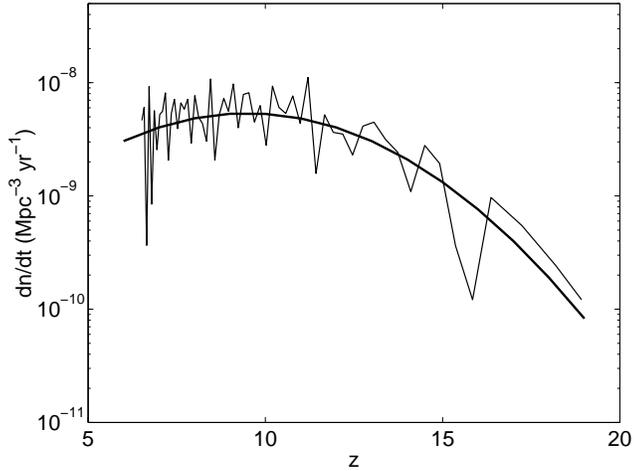}
\caption{The formation rate of 1--2$\times10^8\ M_\odot$ halos per comoving Mpc$^3$ and year, as a function of redshift. The raw simulation data is represented by the thin line, whereas the thick line traces a second-degree polynomial fitted to the data. The halo formation rates used in the analysis are based on the latter.}
\label{fig3}
\end{figure}

According to \citet{Freese et al. 2010}, $10^7 M_\odot$ SMDS form inside $\sim 10^8\ M_\odot$ dark matter halos. This scenario is controversial, since a transition from molecular (mainly H$_2$ and HD) to HI cooling is expected in this halo mass regime, supposedly leading to the formation of entire stellar populations (`first galaxies') rather than single population III stars within a halo. In the standard picture of chemical enrichment in the high-redshift Universe, short-lived population III stars in $\sim 10^5$--$10^6\ M_\odot$ minihalos at $z>15$ moreover pollute the intergalactic medium with metals to such a degree that only a tiny fraction \citep[$< 10^{-3}$ in the models of][]{Stiavelli & Trenti} of the $\geq 10^8 \ M_\odot$ halos that form at $z=10$--15 can be chemically pristine and able to form metal-free SMDS the way \citet{Freese et al. 2010} envision this process. The standard scenario for chemical enrichment can be questioned, however, if a substantial fraction of these first population III in minihalos go through a long-lived dark star phase, since this would delay the production of pair-instability supernovae \citep{Iocco09} and slow down cosmic chemical evolution. For the sake of argument, we put these theoretical problems to the side, and instead attempt to observationally constrain the fraction $f_\mathrm{SMDS}$ of $\sim 10^8\ M_\odot$ CDM halos that form $10^7 M_\odot$ SMDS. If the parameter $f_\mathrm{SMDS}$ is sufficiently low at $z\approx 10$--15, this would explain the lack of such objects in existing survey data. One should keep in mind, however, that both cosmic chemical evolution, the changing intergalactic radiation field and the time-dependent halo merger rate may allow for a $f_\mathrm{SMDS}$ which evolves within this redshift interval.
 
While \citet{Freese et al. 2010} consider SMDS lifetimes $\tau$ of up to $\tau\approx 1$ Gyr, this parameter remains poorly constrained at the current time, and one may also consider values that are several orders of magnitudes smaller. If so, SMDS would only be detectable during a brief period after their formation, hence contributing to the lack of detections in existing data.

In Fig.~\ref{fig3}, we plot the formation rates of 1--$2\times 10^8\ M_\odot$ dark matter halos as a function of redshift, based on high-resolution N-body simulations \citep{Iliev et al. b} of the formation of
high-redshift structures. We use the CubeP$^3$M N-body code\footnote{http://www.cita.utoronto.ca/mediawiki/index.php/CubePM, for a
  description of the code see also \citet{Iliev et al. a}.}
which is based on the particle-mesh  code PMFAST
\citep{Merz et al.} and simulate a cubic comoving volume of size
$6.3\, h^{-1}$~Mpc using $1728^3$ particles of mass
$5.19\times10^3M_\odot$. To identify halos we use a spherical
overdensity halo finder with overdensity parameter fixed to 178 and
a minimum number of particles equal to 20, i.e.\ the minimum halo mass
is $1.04\times10^5M_\odot$. The background cosmology is based on the WMAP 5-year data combined
with constraints from the baryonic acoustic oscillations and
high-redshift supernovae \citep[$\Omega_\mathrm{M} = 0.27, \Omega_\Lambda=0.73, h=0.7, \Omega_b=0.044,
\sigma_8 =0.8, n=0.96$;][]{Komatsu et al.}. 

While \citet{Freese et al. 2010} adopt a formation redshift of $z=15$ for their $10^7 \ M_\odot$ SMDS in $10^8\ M_\odot$ halos, it seems reasonable that such objects should be able to form also in similar halos at slightly lower or higher redshifts, albeit possibly with different formation probabilities $f_\mathrm{SMDS}$. In the following, we present constraints for two different scenarios: A) in  which $10^7 \ M_\odot$ SMDS can form in halos down at least $z\approx 10$ as well, and B) in which the $f_\mathrm{SMDS}$ evolves so strongly with redshift that $10^7 \ M_\odot$ SMDS effectively form only at $z\approx 15$. 

At $z\approx 10$, the formation rate of 1--$2\times 10^8\ M_\odot$ halos is $\mathrm{d}n/\mathrm{d}t\approx 5\times 10^{-9}$ per comoving Mpc$^{3}$ and year (Fig.~\ref{fig3}). This converts into $\approx 580$ halos formed per arcmin$^2$ in the redshift interval $z=9.5$--10.5 where $10^7\ M_\odot$ SMDS would be readily detectable as $J$-band dropouts by \citet{Bouwens et al. a}. This implies that, if the formation of long-lived $10^7\ M_\odot$ SMDS within $10^8\ M_\odot$ halos were a common phenomenon at these redshifts, a survey area like the HUDF (11 arcmin$^2$) would light up like a Christmas tree from the glow of thousands of bright $J$-band dropouts. Since this is clearly not the case, these objects must be exceedingly rare or very short-lived.

\begin{figure}
\includegraphics[width=84mm]{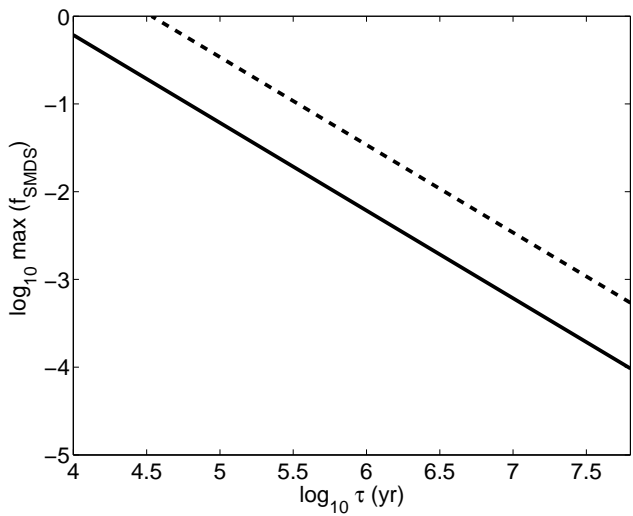}
\caption{Upper limits on the fraction $f_\mathrm{SMDS}$ of 1--$2\times 10^8\ M_\odot$ dark matter halos that form $T_\mathrm{eff}=27000$ K (solid line) and $T_\mathrm{eff}=51000$ K (dashed line) $10^7\ M_\odot$ dark stars at $z\approx 10$, as a function of their lifetimes $\tau$.}
\label{fig4}
\end{figure}

The null detections of $J$-band dropouts in a given survey area can be converted into constraints on $f_\mathrm{SMDS}$ and $\tau$ using the expression:
\begin{equation}
f_\mathrm{SMDS}\leq \frac{\Delta t}{\dot{N}\theta^2\tau},
\end{equation}
where $\dot{N}$ is the number of 1--$2\times 10^8\ M_\odot$ halos forming per unit redshift and arcmin$^{2}$, $\theta^2$ is the angular survey area in arcmin$^2$ and $\Delta t$ is the cosmic age interval per unit redshift. At $z=10$, $\dot{N}\approx580$ and $\Delta t\approx 6.6\times 10^7$ yr. A detailed comparison of the $H_{160}$ fluxes of the $10^7\ M_\odot$ SMDS with the detection thresholds of the surveyed areas reveals that $\theta^2=18.5$ arcmin$^2$ have been imaged to sufficient depth to detect the $T_\mathrm{eff}=27000$ K, $10^7\ M_\odot$ SMDS as $J$-band dropouts, whereas the corresponding area for the $T_\mathrm{eff}=51000$ K, $10^7\ M_\odot$ SMDS is $\theta^2=3.3$ arcmin$^2$. From this we derive the two sets of upper limits on $f_\mathrm{SMDS}$ as a function of $\tau$ included in Fig.~\ref{fig4}. 

The resulting constraints are strong: e.g. $\log_{10} f_\mathrm{SMDS}\leq -3.2\ (-2.5)$ if $\tau\sim 10^7$ yr and $\log_{10} f_\mathrm{SMDS}\leq -2.2\ (-1.5)$ if $\tau\sim 10^6$ yr for the $T_\mathrm{eff}=27000$ (51000) K SMDS. These upper limits formally apply only to the value of $f_\mathrm{SMDS}$ attained at $z=10\pm 0.5$. Lifetimes in excess of $\Delta t = 6.6\times 10^7$ yr would make the constraints even stronger, but would require additional assumptions concerning the redshift evolution of $f_\mathrm{SMDS}$. 

In scenario B, where $f_\mathrm{SMDS}$ is assumed to be effectively zero at $z=10$, the existing data can still be used to set upper limits on $f_\mathrm{SMDS}$ at $z=15$ \citep[the formation redshift assumed by][]{Freese et al. 2010}, provided that the SMDS forming at $z=15$ have sufficiently long lifetimes to survive until $z=10$. In the adopted cosmology, this requires $\tau > 2.1\times 10^8$ yr. By adopting a formation rate of $\mathrm{d}n/\mathrm{d}t\approx 1\times 10^{-9}$ per comoving Mpc$^{3}$ and year for 1--$2\times 10^8\ M_\odot$ halos (see Fig.~\ref{fig3}), we arrive at $\log_{10 }f_\mathrm{SMDS}\leq-2.9\ (-2.2)$ for the $T_\mathrm{eff}=27000$ (51000) K, $10^7\ M_\odot$ SMDS from \citet{Freese et al. 2010}.

\section{Discussion}
\label{discussion}
As demonstrated by \citet{Freese et al. 2010}, {\it JWST} can detect $\sim 10^5\ M_\odot$ SMDS out to $z\approx 10$ and $\sim 10^7\ M_\odot$ SMDS out to $z\approx 15$. However, the constraints already placed on $\sim 10^7\ M_\odot$ SMDS by existing data imply that the prospects of detecting objects in this mass range with the {\it JWST} may be rather bleak. At $z=15$ we predict $\dot{N}\approx 30$ halos with mass 1--$2\times 10^8 M_\odot$ forming per arcmin$^2$ and unit redshift. With $\Delta t\approx2.6\times10^7$ yr and a field of view covering $\theta^2=4.84$ arcmin$^2$, a single {\it JWST} detection of a $\sim 10^7\ M_\odot$ SMDS at $z=15$ would suggest $\log_{10} f_\mathrm{SMDS}\approx -1.8$ if $\tau=10^7$ yr. However, this combination of $f_\mathrm{SMDS}$ and $\tau$ has already been ruled out at $z=10$ (Fig.~\ref{fig4}). Hence, if $f_\mathrm{SMDS}$ and $\tau$ are approximately the same at $z=15$ and $z=10$ (as in our scenario A), our constraints predict that no $10^7\ M_\odot$ SMDS will be detectable within a single {\it JWST} field at $z=15$. Of course, {\it JWST} observations would still be highly relevant for dark stars at lower masses. As demonstrated in Fig.~\ref{fig1}, ${\it HST}$ observations through lensing clusters may also be able to set constraints on SMDS down to masses of $\sim 10^5\ M_\odot$.

\section{acknowledgements}
E.Z, C-E.R and G.\"O. acknowledge funding from the Swedish National Space Board, and E.Z., P.S., C-E.R, S.S., G.\"O. and G.M. from the Swedish Research Council. F.~I. is supported by European Community research program FP7/2007/2013 within convention \#235878.  PRS is supported in part by grants NSF AST 0708176, NASA NNX07AH09G, Chandra Grant SAO TM8-9009X, the Texas Advanced Computing Center (TACC) at UT Austin, and NSF TeraGrid grants TG-AST0900005 and TG-080028N. The authors are indepted to Cosmin Ilie and Katherine Freese for bringing a numerical mistake in the published manuscript to our attention. This error has been corrected in the present version of the paper.

\end{document}